%% file: comm.tex
\documentclass[11pt,epsf]{article}

\usepackage{array}
\usepackage{graphicx}

\usepackage{multirow}
\usepackage{caption}

\begin{document}

\input pp1-4.tex

\newpage
\renewcommand{\baselinestretch}{1}
\input figures1.tex

\end{document}

%% file: pp1-4.tex




\title{Relations between $\beta$ and $\delta$ for QP and LP\\ in Compressed Sensing Computations}

\author{Jun Zhang, Jun Wang, and Guangwu Xu
\\Department of Electrical Engineering and Computer Science
\\University of Wisconsin-Milwaukee\\
Milwaukee, WI 53211}

\date{}
\maketitle{}

\section*{Abstract}

In many compressed sensing applications, linear programming (LP) has been used to reconstruct a  sparse signal. When observation is noisy, the LP formulation is extended to allow an inequality constraint and the solution is dependent on a parameter $\delta$, related to the observation noise level. Recently, some researchers also considered quadratic programming (QP) for compressed sensing signal reconstruction and the solution in this case is dependent on a Lagrange multiplier $\beta$. In this work, we investigated the relation between $\delta$ and $\beta$ and derived an upper and a lower bound on $\beta$ in terms of $\delta$. For a given $\delta$, these bounds can be used to approximate $\beta$. Since $\delta$ is a physically related quantity and easy to determine for an application while there is no easy way in general to determine $\beta$, our results can be used to set $\beta$ when the QP is used for compressed sensing. Our results and experimental verification also provide some insight into the solutions generated by compressed sensing.

\section*{1. Introduction}

In many compressed sensing applications, signal reconstruction is
carried out by solving a linear programming (LP) problem
\cite{Can05}. When observation is noiseless, the LP problem is:

\begin{equation}
\mbox{(LP)}: \quad \mbox{min} ||u||_1 \qquad \mbox{subject to $Au =
b$}
\end{equation}
where $u$ is the signal to be reconstructed ($n$-dim vector, sufficiently sparse), $A$ is an  $m\times n$ sampling matrix, with $m \ll n$, $b$ is the observation (an $m$-dim vector), and the minimization is with respect to $u$. When the observation is noisy (with bounded noise), the LP formulation is extended to

\begin{equation}
\mbox{(LPn)}: \quad \mbox{min} ||u||_1 \qquad \mbox{subject to $||Au -
b||^2_2 \le \delta^2$} \label{LPp}
\end{equation}
where $\delta^2$ is a bound on noise power, or {\it noise level}. Although strictly speaking  this is no longer a linear programming problem (it is still a convex optimization problem), because its relation to eqn (\ref{LP}) and for the sake of simplicity, we will still refer to it as a part of the LP problem (LPn).

In some work, a quadratic programming (QP) problem, with

 \begin{equation}
\mbox{(QP)}: \quad \mbox{min}  \bigg\{\frac{1}{2}||Au - b||_2^2 + \beta ||u||_1 \bigg\}
\label{QP}
\end{equation}
has been considered for compressed sensing. For example, Fuchs \cite{Fuc04, Fuc05} and Troop \cite{Tro06} used the QP formulation to study the theoretical properties of the solutions of various  compressed sensing problems. Similarly, Chen et al. (e.g., see \cite{Che08}) used a QP-type formulation in several papers for compressed sensing based 3D CT (computer tomography) [in their case, the $l_1$ norm is replaced by total variation (TV)]. Finally, the QP problem is closely related to (in fact equivalent to) the Lasso procedure \cite{Las} which is widely used in statistics, pattern recognition, and data mining.

Given the LP and QP formulations, a natural question is: when are they
the same, i.e., producing the same results? When the observation is
noiseless, Fuchs \cite{Fuc04} showed that the QP becomes the same as
the LP when $\beta \rightarrow 0^+$ (i.e., from the right).  When
the observation is noisy, Fuchs \cite{Fuc05} pointed out that for a
given noise bound $\delta^2$ in the LPn, there exists a $\beta >0$
for the QP such that the resulting QP is the same as the LPn, i.e.,
they produce the same solutions; in fact, this can be established through
the theory of duality \cite{Boy04}. However, Fuchs also mentioned
that it is difficult to find an explicit (e.g., an analytic) relation
between this $\beta$ and the $\delta^2$ in LPn. This means that when
a QP algorithm is used in practice for compressed sensing, such as
in the previously mentioned 3D CT and Lasso applications, it may not
actually be performing compressed sensing (as defined by the LPn)
since, in these applications, it is unclear how to find the
``right'' $\beta$. Indeed, in practice $\beta$ is often selected
experimentally. However, as illustrated in Fig. 1 (figures are all at the end of the paper), for a given compressed sensing problem with a given noise level ($\delta^2$), the solution from the QP is dependent on parameter $\beta$ and most of the time, it is not the same as that from the LPn (this is illustrated through the QP's reconstruction error for various $\beta$ values).

In this paper, we attempt to find an analytic relation between
$\beta$ and $\delta^2$ (or equivalently, between $\beta$ and
$\delta$). Such a relation is useful in three respects. First, it may
allow us to gain more insight into the relation between the LP and
the QP problems, thereby more insight into the nature of the
solution of the compressed sensing problem. Second, if in practice
we want to use the QP or Lasso for some reason (e.g., familiarity, easier
implementation, or faster speed) as an algorithm for compressed
sensing, we can obtain or estimate the appropriate $\beta$ from
$\delta$, which, as the noise level, is a physical parameter and
usually readily available. Finally, many signal/image processing and computer vision problems are solved by a Bayesian formulation where an energy function related to the posterior probability distribution is minimized. Usually, the energy function is a sum of two terms: the first is related to the observation model and is similar to, or the same as, the first term in QP [see eqn (\ref{QP})] and the second is related to prior constraints and is similar to the second term in the QP. The two terms are ``balanced" by a parameter $\beta$ just like that in the QP formulation. In many such Bayesian applications, selecting the value of $\beta$ is a problem and there is no analytic/theoretical guidance. Our work sheds light on this problem and could potentially be used to find a solution.

The rest of the paper is organized as
follows. In Section 2, we derive some analytic relations between
$\beta$ and $\delta$ and in Section 3, we verify and illustrate some
of these relations experimentally. Finally, in Section 4, we provide
conclusions.

\section*{2. Analytic Results}

In this section, we derive two relations between $\beta$ and
$\delta$, one in inequalities and the other in an equality.

\subsection*{2.1. Inequality Relations}

Suppose for a noise level or bound $\delta^2$, the LPn problem of
eqn (\ref{LPp}) has a sparse solution. Then, as described in Section
1, there is a $\beta > 0$ such that for this $\beta$, the solution
of the QP problem of (\ref{QP}) is the same as that of the LPn. Let
$x$ be this ``common'' solution and suppose it has $k$ non-zero
components. According to Fuchs \cite{Fuc04, Fuc05}, this solution
must satisfy the following condition (can also be viewed as part of
the KKT condition \cite{Boy04})

\begin{equation}
{\bar A}^T (b - {\bar A}{\bar x}) = \beta \mbox{sgn}({\bar x})
\label{cond-1}
\end{equation}
where $\bar x$ is the ``reduced solution vector,'' made up by the non-zero components of $x$, $\bar A$ is a $m\times k$ matrix made up by the columns of matrix $A$ corresponding to $\bar x$, and $\mbox{sgn}(\cdot)$ is the usual sign function that for a scalar $t$

\begin{equation}
\mbox{sgn}(t) = \left\{ \begin{array}{ll}
-1, & \mbox{if $t<0$} \\
+1, & \mbox{if $t>0$}
\end{array} \right.
\end{equation}
while for a vector $v$, $\mbox{sgn}(v)$ is applied component-by-component, leading to a vector of $+1$s and $-1$s.

Now, taking the $l_2$ norm square on both sides of (\ref{cond-1}),
we have

\begin{equation}
(b - {\bar A}{\bar x})^T ({\bar A}{\bar A}^T) (b - {\bar A}{\bar x})
= \beta^2 || \mbox{sgn}({\bar x}) ||^2
= \beta^2 k
\label{basic-1}
\end{equation}
where we used the fact that

\begin{equation}
||\mbox{sgn}(\bar x)||^2 = (\sqrt{k})^2 = k
\end{equation}

Note that ${\bar A}{\bar A}^T$ is a $m\times m$ correlation matrix and is semi
positive definite. Hence, its eigenvalues are non-negative. Denoting the largest among these as $\lambda_{max}$ and using the relation between matrix norm and maximum eigenvalues  \cite{Hor85}, we have

\begin{equation}
(b - {\bar A}{\bar x})^T ({\bar A}{\bar A}^T) (b - {\bar A}{\bar x}) \le
\lambda_{max} || b - {\bar A}{\bar x}||^2
\end{equation}
Because of eqn (\ref{basic-1}), we can also write this as

\begin{equation}
\beta^2 k \le
\lambda_{max} || b - {\bar A}{\bar x}||^2
\end{equation}

As described previously, $\bar x$ (and $x$) is also a solution of
the LPn problem, it satisfies the inequality constraint of eqn
(\ref{LPp}). In fact, based on the results in the Appendix, this
solution achieves equality

\begin{equation}
|| b - {\bar A}{\bar x}||^2 = \delta^2
\label{maxp}
\end{equation}
Hence, we have
\begin{equation}
\beta^2 k \le
\lambda_{max} \delta^2
\end{equation}
That is,

\begin{equation}
\beta \le
\sqrt{\frac{\lambda_{max}}{k}} \delta
\label{basic-2}
\end{equation}
Given $\delta$, this provides an upper bound on $\beta$. In practice, we could also use this upper bound as an approximation to $\beta$ with

\begin{equation}
\beta \simeq \sqrt{\frac{\lambda_{max}}{k}} \delta
\end{equation}
As demonstrated in Section 3, this approximation can often be quite good. Finally, it is interesting to note that if we let $\delta
\rightarrow 0^+$, the LPn problem becomes the LP problem. In this
case, the upper bound suggests that $\beta \rightarrow 0^{+}$, reproducing
Fuch's noiseless result for the relation between $\beta$ and
$\delta$ (see Section 1).

Using the techniques for deriving the above upper bound, we can also derive a lower bound. However, since ${\bar A}{\bar A}^T$ is semi positive definite rather than positive definite, this is slightly less straightforward and requires some approximations. Specifically, since $\bar A$ is an $m\times k$ matrix and since generally, $m > k$ (in practice, $m$ is usually on the order of $5k$ \cite{Can05}), the rank of matrix ${\bar A}{\bar A}^T$ is at most $k$. Since ${\bar A}{\bar A}^T$ is an $m \times m$ matrix, it has at most $k$ non-zero eigenvalues. Assume this to be the case and denote the smallest non-zero eigenvalue be denoted as $\lambda_{min}$. Let the eigenvectors of ${\bar A}{\bar A}^T$ be $e_1, e_2, \ldots, e_m$, where they are ordered according to the value of their eigenvalues, $e_1$ for $\lambda_{max}$, $e_k$ for $\lambda_{min}$, and $e_{k+1}, \ldots, e_m$ for 0 eigenvalue. Since $b - {\bar A}{\bar x}$ is an $m$-dimensional vector, it can be represented by $e_1, e_2, \ldots, e_m$, with

\begin{equation}
b - {\bar A}{\bar x} = \sum_{i=1}^m \alpha_i e_i
\end{equation}
where $\alpha_i$ are representation coefficients. From this, we have

\begin{equation}
(b - {\bar A}{\bar x})^T ({\bar A}{\bar A}^T) (b - {\bar A}{\bar x})
= \bigg(\sum_{i=1}^m \alpha_i e_i\bigg)^T  ({\bar A}{\bar A}^T)
\bigg(\sum_{i=1}^m \alpha_i e_i\bigg)
= \sum_{i=1}^m \lambda_i\alpha_i^2  = \sum_{i=1}^k \lambda_i \alpha_i^2
\ge \lambda_{min}  \sum_{i=1}^k \alpha_i^2.
\label{basic-3}
\end{equation}

Now, we find an estimate of $\sum_{i=1}^k \alpha_i^2$. First we notice that the larger sum

\begin{equation}
\sum_{i=1}^m \alpha_i^2 = ||b - {\bar A}{\bar x}||^2.
\end{equation}
Hence, from eqn (\ref{maxp}) we have

\begin{equation}
\sum_{i=1}^m \alpha_i^2 = \delta^2
\end{equation}
Furthermore, we notice that $b - {\bar A}{\bar x} = b - Ax = w$ is the noise. Assume the noise is white (i.e., uncorrelated), on average $\alpha_i^2$ are roughly the same \cite{Van68}, at $\delta^2/m$. Hence, we have

\begin{equation}
\sum_{i=1}^k \alpha_i^2 \simeq \frac{k}{m} \delta^2
\label{basic-4}
\end{equation}

Now, combine eqns (\ref{basic-1}), (\ref{basic-3}), and (\ref{basic-4}), we have
\begin{equation}
\lambda_{min} \frac{k}{m} \delta^2 \le \beta^2 k
\end{equation}
That is,
\begin{equation}
\sqrt{\frac{\lambda_{min}}{m}} \delta \le \beta
\label{basic-5}
\end{equation}
This provides a lower bound on $\beta$ for a given $\delta$.

The only problem left now is to find the eigenvalues $\lambda_{min}$
and $\lambda_{max}$. In general these eigenvalues are dependent on
the specifics of the $A$ matrix, such as which columns correspond to
the non-zero elements of $x$. However, there is an important case of
practical importance where these eigenvalues can be found relatively
easily. This is the widely used case where $A$ is an i.i.d. Gaussian
random matrix. In this case, it has been shown in previous work
\cite{Gem80, Can05} that the smallest and the largest eigenvalues
for ${\bar A}{\bar A}^T$ are asymptotically

\begin{equation}
\lambda_{min} \simeq m\sigma^2(1 - \sqrt{\gamma})^2, \qquad
\lambda_{max} \simeq m\sigma^2(1 + \sqrt{\gamma})^2,
\end{equation}
where $m$ is the number of rows in $A$ and $\bar A$, $\sigma^2$ is the variance of each component of $A$ (and $\bar A$), $\gamma = k/m$ (recall that $k$ is the dimension of $\bar x$, also the number of columns of $\bar A$). Plugging these into the bounds of eqns (\ref{basic-2}) and (\ref{basic-5}), we then have

\begin{equation}
(1 - \sqrt{\gamma})\sigma\delta \le \beta \le
\frac{(1 + \sqrt{\gamma})\sigma}{\sqrt{k/m}}\delta
\label{ineq}
\end{equation}
When a compressed sensing application uses a Gaussian random
sampling matrix, the inequality of (\ref{ineq}) can be used to find
the range of, or estimate, $\beta$ from a given $\delta^2$.

\subsection*{2.2. An Equality Relation}

If we add an additional assumption to the derivations in Section
2.1, we can obtain an equality relation between $\delta$ and
$\beta$. Specifically, if we assume that in eqn (\ref{cond-1}) the $k\times k$ matrix
${\bar A}^T{\bar A}$ is invertible, as Fuchs did in his papers
\cite{Fuc04, Fuc05},  then eqn (\ref{cond-1}) becomes

\begin{equation}
\bar{x} = (\bar{A}^T \bar{A})^{-1}(\bar{A}^T b -\beta \mbox{sgn$(\bar{x})$})
\label{QPS}
\end{equation}
and this provides a solution to the QP problem. As we mentioned
previously, when $\beta$ matches $\delta$, this solution is the same as
that of the LPn. Furthermore, it can be shown that the solution of
the LPn must satisfy the inequality with equality (see Appendix).
Hence, the solution of (\ref{QPS}) should satisfy

\begin{equation}
||\bar{A}\bar{x} - b||^2_2 = \delta^2
\end{equation}
or
\begin{equation}
\bar{x}^T \bar{A}^T \bar{A} \bar{x} -2 b^T \bar{A}\bar{x} + ||b||^2
= \delta^2 \label{s2}
\end{equation}
where we have dropped the subscript 2.

Plugging the right hand side of (\ref{QPS}) into (\ref{s2}) and
denoting $\mbox{sgn}(\bar{x})$ as vector $c$ to simplify notation, the
first term of (\ref{s2}) becomes

\begin{eqnarray}
&& \bar{x}^T\bar{A}^T \bar{A} \bar{x} = [(\bar{A}^T
\bar{A})^{-1}(\bar{A}^T b  -\beta c)]^T
\bar{A}^T \bar{A} [(\bar{A}^T \bar{A})^{-1}(\bar{A}^T b -\beta c)] \nonumber \\
&& = (\bar{A}^T b  -\beta c)^T [(\bar{A}^T \bar{A})^{-1}]^T
(\bar{A}^T \bar{A}) [(\bar{A}^T \bar{A})^{-1}(\bar{A}^T b -\beta c)]\nonumber\\
&& = [(\bar{A}^T b)^T  -\beta c^T] [(\bar{A}^T \bar{A})^{-1}(\bar{A}^T b -\beta c)] \nonumber \\
&& = b^T \bar{A} (\bar{A}^T \bar{A})^{-1} \bar{A}^T b
+ c^T (\bar{A}^T \bar{A})^{-1} c \beta^2
-2 b^T \bar{A} (\bar{A}^T \bar{A})^{-1} c \beta
\label{Qterm}
\end{eqnarray}
where we have used the fact that $\bar{A}^T\bar{A}$ is symmetrical and so is its inverse, i.e., $[(\bar{A}^T\bar{A})^{-1}]^T = (\bar{A}^T\bar{A})^{-1}$.

Similarly, for the second term of (\ref{s2}), we have
\begin{eqnarray}
&& -2 b^T \bar{A}\bar{x} =
- 2 b^T \bar{A} [(\bar{A}^T \bar{A})^{-1}(\bar{A}^T b -\beta c)] \nonumber \\
&& = - 2 b^T \bar{A}(\bar{A}^T \bar{A})^{-1} \bar{A}^T b
+ 2 b^T \bar{A}(\bar{A}^T \bar{A})^{-1} c \beta
\label{Lterm}
\end{eqnarray}
Combine this and the results of (\ref{Qterm}) into (\ref{s2}), we
have

\begin{equation}
||\bar{A}\bar{x} - b||^2 \nonumber \\
 = - b^T \bar{A}(\bar{A}^T \bar{A})^{-1} \bar{A}^T b
+  c^T (\bar{A}^T \bar{A})^{-1} c \beta^2 + ||b||^2 = \delta^2
\end{equation}
where the terms linear in $\beta$ (\ref{Qterm}) and (\ref{Lterm})
canceled each other. From this, we can find $\beta$ in terms of
$\delta$ as:

\begin{equation}
\beta = \sqrt{\frac{\delta^2 - ||b||^2 + b^T \bar{A}(\bar{A}^T \bar{A})^{-1} \bar{A}^T b}{c^T (\bar{A}^T \bar{A})^{-1} c}}
\end{equation}
Although this equality provides a more explicit relation between
$\beta$ and $\delta^2$, in practice it is more difficult to use than the inequalities of Section 2.1 since $c$ and $\bar{A}$ are generally not known before QP and LP
are performed.

\section*{3. Experimental Verification}

In this section, we provide some experimental (simulation) results
that verify and illustrate the inequality relations between $\delta$
and $\beta$ derived in Section 2.1. In each experiment, we picked a
LPn problem with a given noise level $\delta^2$  [see eqn
(\ref{LPp})] and formed a corresponding QP problem [see eqn
(\ref{QP})]. Then, the QP problem was solved for a range of $\beta$s
and from the resulting solutions, we would try to identify the best
$\beta$ corresponding to the $\delta$ since, according to the theory
of duality, the best $\beta$ should result in the same solution as that of
the LPn with $\delta^2$. From this, we could see if, or how well, the
best $\beta$ satisfies the upper and lower bounds derived in Section
2.1. Next, we describe the specific steps in our experiments.

\subsection*{3.1. Experiment Steps}

Each experiment consists of the following steps:

\begin{enumerate}

\item  Generate a sparse random $n$-dimensional signal $x^*$.

\item Generate a noisy observed signal $b = Ax^* + w$, where $A$ is an $m\times
n$ Gaussian random sampling matrix with $m < n$ and component variance $\sigma^2$ and $w$ is an $m$-dimensional additive white Gaussian noise vector with variance
$\sigma_w^2$.

\item Reconstruct $x^*$ by solving the LPn problem with $\delta^2$ set to
$\delta^2 = m\sigma_w^2$. Denote the resulting solution as ${\hat
x}(\delta)$.

\item Reconstruct $x^*$ by solving the QP problem with a range of $\beta$s.
Denote the resulting solution as ${\hat x}(\beta)$.

\item Compare the minimum obtained by LPn, i.e., $||{\hat x}(\delta)||$, and the maximum
of the dual function obtained by QP, $g(\beta)$ (more details later).

\item Compare the normalized reconstruction errors (i.e., $||x^* - {\hat x}||_2/||x^*||$) obtained by LPn and QP (${\hat x}$ could be either ${\hat x}(\delta)$ or ${\hat x}(\beta)$).

\item Find ``the best $\beta$" (that produces the same QP solution as the LPn solution) and compare this $\beta$ with the bounds in eqns (\ref{basic-2}) and (\ref{basic-5}).

\end{enumerate}

We now explain each of these steps in some detail. In Step 1, the original random sparse signal $x^*$ was obtained from examples
provided/generated by the L1 Magic software \cite{L1M} (which
uses these to illustrate the workings of compressed sensing algorithms). Specifically, in our experiments $x^*$ is a sparse random vector
of dimension $n=256$ and its non-zero components consists of $k=24$
randomly placed +1s and -1s (also randomly chosen), as shown in Fig.
2.

In Step 2, the noisy observed signal $b$ was generated using a
$m\times n$ Gaussian random sampling matrix $A$ with $m=100$ and
component variance $\sigma^2 = 1$; for the additive noise $w$, we
used a white Gaussian noise with variance $\sigma^2_w$ (whose value
is different in different experiments, more details later). Some
typical noisy observed signals are also shown in Fig. 2.

In Step 3, the LPn problem was solved with a log barrier algorithm
in L1 Magic and in Step 4, the QP problem was solved with the L1
Regularization software developed by Kim et al \cite{L1R}.

In Step 5, what we are really doing is to use results of the duality
theory \cite{Boy04} to find the best $\beta$. Specifically, for the
LPn problem, one can define a dual function

\begin{equation}
g(\lambda) = \mbox{inf}_u \bigg\{ ||u||_1 + \lambda \big(||Au -
b||_2^2 -\delta^2\big)\bigg\}
\end{equation}
Because the LPn satisfies the strong duality condition (see
\cite{Boy04}), we have

\begin{equation}
||{\hat x}||_1 \ge g(\lambda) \quad \mbox{for all $\lambda \ge 0$}
\end{equation}
where $\hat x$ is the solution of the LPn problem and equality is achieved at the best $\lambda$, denoted as $\lambda_0$, with
\begin{equation}
\lambda_0 = \arg\max g(\lambda) \label{SD1} \quad \mbox{and} \quad ||{\hat x}||_1 = g(\lambda_0)
\end{equation}
Now, the dual function $g(\lambda)$ can be linked to the QP solution: we can re-write it as
\begin{equation}
g(\lambda) = \mbox{inf}_u \bigg\{ ||u||_1 + \lambda
\bigg(||Au - b||_2^2 -\delta^2\bigg)\bigg\} = 2 \lambda \mbox{inf}_u
\bigg\{\frac{1}{2}||Au - b||_2^2 + \frac{1}{2\lambda}||u||_1 \bigg\}
-\lambda\delta^2
\end{equation}
where the $\mbox{inf}_x \{\cdot\}$ is the QP solution with $1/2\lambda = \beta$ . In this way, whenever a QP problem with $\beta$ is solved, we can compute a corresponding $g(\lambda)$ with $\lambda = 1/2 \beta$. In this sense, we can write (re-parameterize) $g(\lambda)$ as $g(\beta)$ and the strong duality of eqn (\ref{SD1}) can be re-written in terms of $\beta$ as

\begin{equation}
\beta_0 = \arg\max g(\beta) \label{SD2} \quad \mbox{and} \quad ||{\hat x}||_1 = g(\beta_0)
\label{best-beta}
\end{equation}
where $\beta_0$ is the best $\beta$ (which makes the QP having the
same solution as that of the LP).

Finally, Steps 6 and 7 are relatively straightforward and we will discuss our experimental results next.

\subsection*{3.2. Experimental Results}

Some typical experimental results are shown in Figs. 3-14. In Fig.
3, the observation noise variance is $\sigma_w^2 = 0.0225$,
corresponding to an SNR of 30dB (low noise) and a
noise bound of $\delta^2 = m\sigma_w^2 = 100\times 0.0225 = 2.25$. Fig. 3 contains
information obtained in Step 5, i.e., the minimum achieved by the
LPn ($||{\hat x}||_1$), the re-parametized dual function $g(\beta)$,
and the upper and lower bounds for $\beta$ we derived in Section
2.1. Note that since the minimum achieved by PLn is a number
(constant) while the dual function is a function of $\beta$, we
presented the former as a constant line. Similarly, the bounds are
numbers, i.e., specific values of $\beta$, hence are presented as vertical lines.

From the results of Fig. 3, we can make two observations. First, the
experimental results agrees with the prediction of the duality
theory. That is, the $g(\beta)$ curve is always below the $||{\hat x}||_1$ line
and for the ``best" $\beta$, the curve approaches the line. Second, the
best $\beta$ falls in an interval predicted/defined by our upper and
lower bounds  (almost right in the middle of the interval). This is very encouraging.

Fig. 4 compares the normalized reconstruction errors (see Step 6)
for the LPn and QP. The former is a number, presented as a
horizontal line while the latter is a function of $\beta$, hence is
a curve. At the ``best" $\beta$ [i.e., when eqn (\ref{best-beta}) is satisfied or, when the curve meets the straight line in Fig. 3], the QP and LP have the same reconstruction errors.
Hence, looking at reconstruction errors of QP and LP provides another potential way\footnote{Sometimes, the normalized QP error curve can intersect the LP error line at more than one $\beta$, in this case, we need to rely on the dual function to identify the best $\beta$.} to identifying the ``best" $\beta$, as can be seen in Fig. 4; for the ``best" $\beta$, the QP has the same reconstruction error as that of LP. From Fig. 4, we can
also observe that $\beta$ can have a strong effect on reconstruction
error. Finally, we note that the best $\beta$ does not necessarily
lead to minimum reconstruction error for QP since the best $\beta$
is best in the sense of duality theory (providing the same solution
as that of LPn), not in the sense of minimum reconstruction error.
Currently, it is not obvious as to how to find the best $\beta$ in this latter sense.

To ensure that our results in Figs. 3 and 4 are no accidents, we
repeated that experiment 100 times (each time with a new random
sparse signal, random sampling matrix, and additive noise vector)
and averaged their results. These are shown in Figs 5 and 6. As can
be seen, the nature of the results are the same as that of Fig. 3
and 4.

Finally, we repeated the experiment for Figs. 3-6 for a higher noise
level, with $\sigma_w^2 =0.2025$, corresponding to an SNR of 20dB
(heavy noise) and a noise level of $\delta^2 = 100\times 0.2025 = 20.25$. The results are
presented in Figs. 7-10. The nature of the results is the same as
that of Figs. 3-6: our derived bounds worked well. Furthermore,
compared with Figs 3-6, we can see that as the noise reduces (from
20dB to 30dB SNR), the bounds and the best $\beta$ move to the left, agreeing with the theoretical prediction that the best $\beta
\rightarrow 0^+$ when the noise level reduces to 0.

\section*{Appendix}


Consider the solution to the $\ell_1$ minimization problem
\[
(P_1) \quad \min\|u\|_1 \quad \mbox{ subject to } \quad  \|Au - b\|_2 \le \delta.
\]

First we note that if $\|b\|_2 \le \delta$, then $\hat x=0$ is the
solution to $(P_1)$. To avoid this trivial case, we assume $\|b\|_2
> \delta$.

The following result belongs to a well-known result in convex optimization,
known as the maximum principle. We include it here for the reader's
convenience. Plus, our proof is specialized to the $(P_1)$ problem.

\vspace{.1in}
\noindent {\bf Maximum Principle:} Let $\hat x$ be a minimizer of $(P_1)$ and if $\hat x\neq 0$, then
\[
\|A \hat x - b\|_2 = \delta.
\]

\noindent
{\it Proof}: In fact, since $\hat x\neq 0$, we may assume $\hat x(i_0)\neq 0$ for
some $i_0$ ($x(i)$ is the $i$th component of $x$). Suppose $\hat x$
is not on the boundary, then
\[
d=\delta-\|A \hat x - b\|_2 > 0.
\]
Choose a small $t$ such that $|\hat x(i_0)-t|< |\hat x(i_0)|$ and $\|A
(\hat x-\hat x')\|_2< d$ ( here $\hat x'(i)=\hat x(i)$ for $i\neq
i_0$ and $\hat x'(i_0)=\hat x(i_0)-t$), we get a contradiction
because $\|A \hat x' - b\|_2<\delta$ and $\|\hat x'\|_1 < \|\hat
x\|_1$. This proves the maximum principle.

%% file: figures1.tex
%





\begin{figure}[htp]
\centering
\includegraphics[scale=0.5]{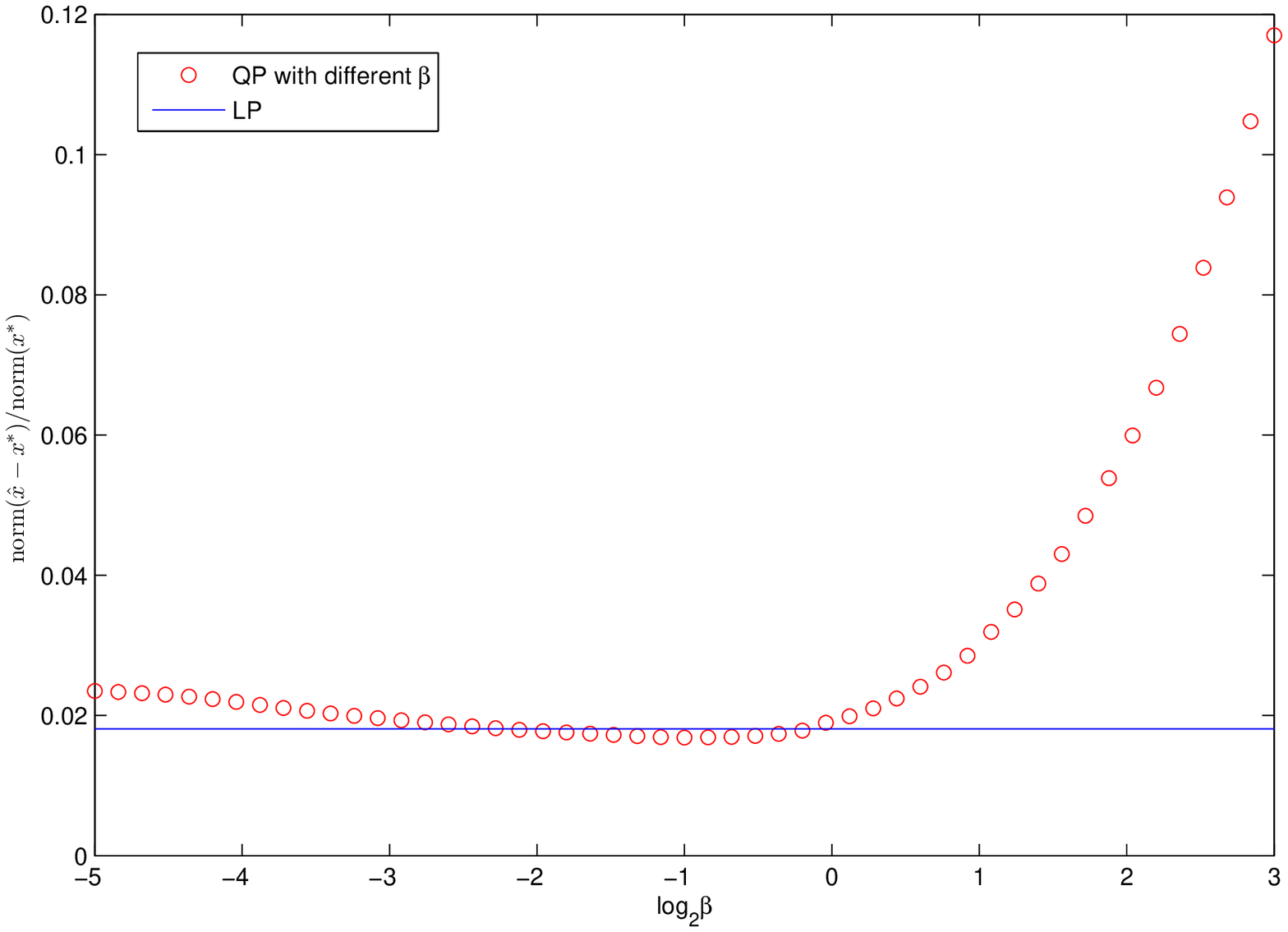}
\caption{QP's Reconstruction Results as a Function of $\beta$. Curve: QP's reconstruction error as a function of $\beta$; Straight line: LP's reconstruction error. $n=100, k=10,m=50,\delta^2 = 0.75$ (SNR = 37dB)}\label{fig:fig1}

\includegraphics[scale=0.5]{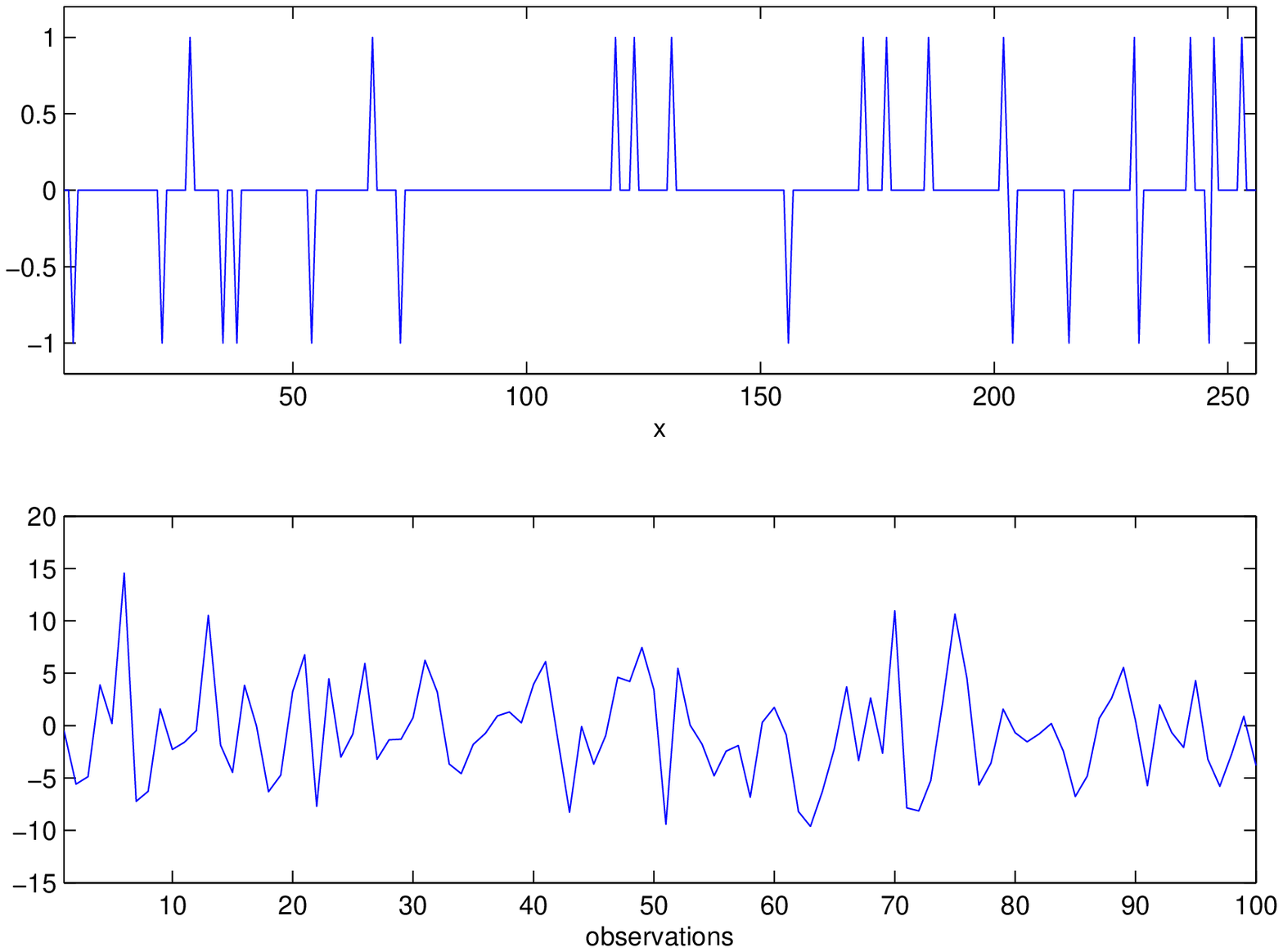}
\caption{A Sparse Random Signal and its Noisy Observation. Top: the sparse random signal, bottom: the noisy observed signal with an SNR of 20dB.}\label{fig:fig2}
\end{figure}

\begin{figure}[htp]
\centering
\includegraphics[scale=0.5]{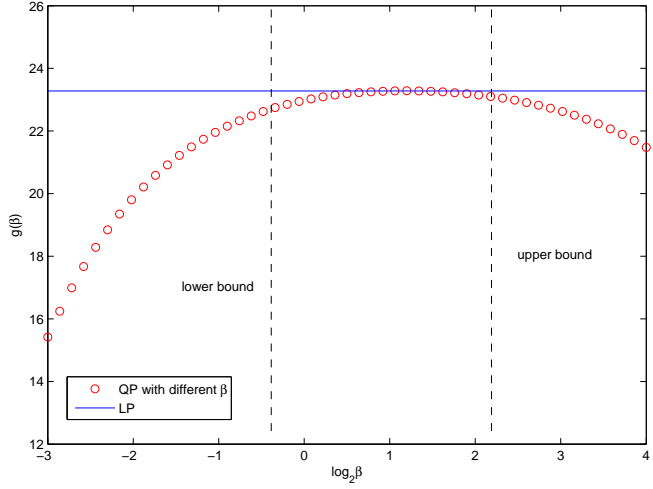}
\caption{The Minimum from LP and the Dual Function $g(\beta)$ from QP: the 30dB SNR Case.
Straight line: the minimum from LP, curve: the dual function from QP.}\label{fig:fig3}
\end{figure}

\begin{figure}[htp]
\centering
\includegraphics[scale=0.5]{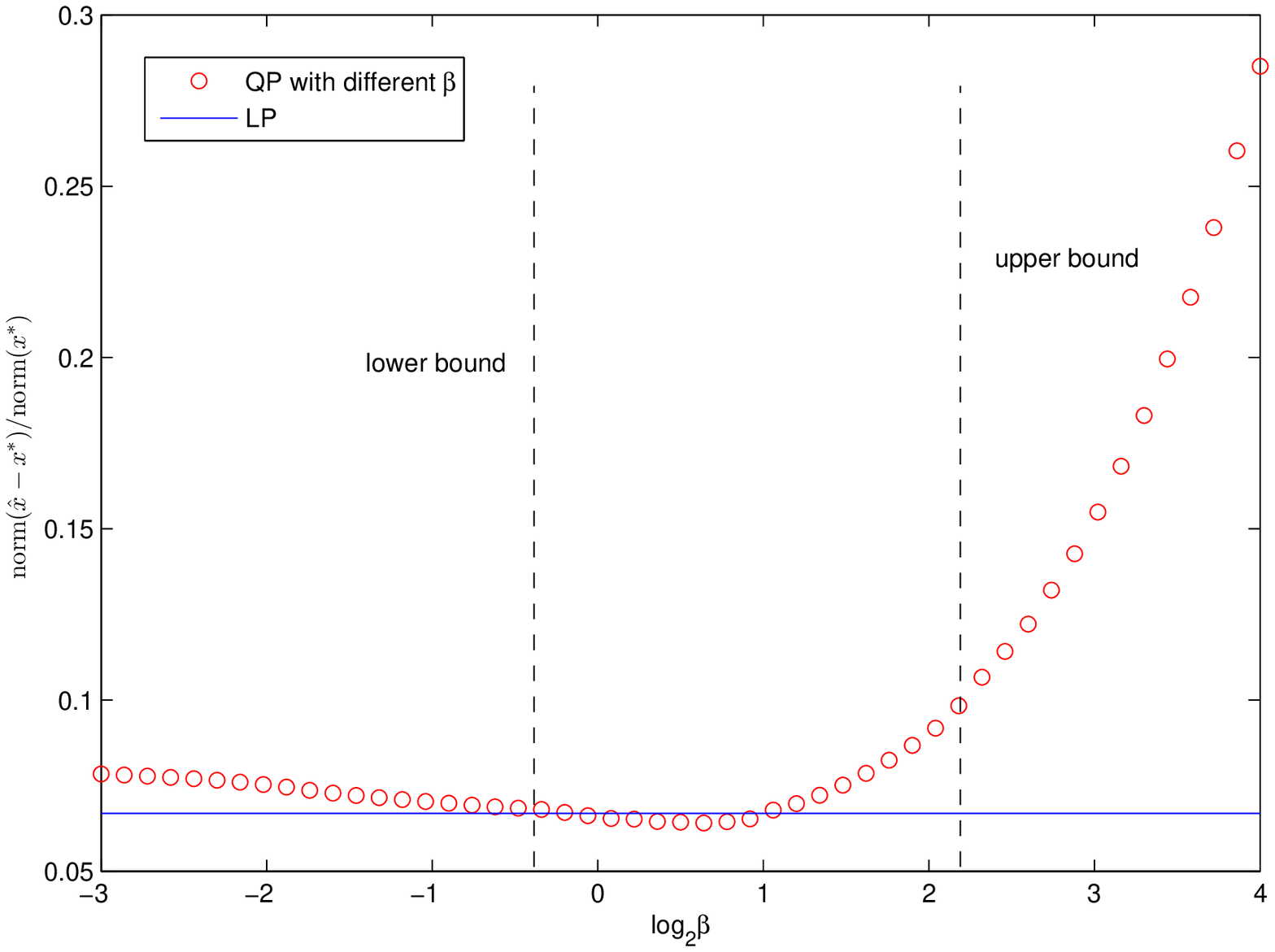}
\caption{Recovery Error from LP and QP: the 30dB SNR Case. The straight line: LP, the curve: QP.}\label{fig:fig4}
\end{figure}

\begin{figure}[htp]
\centering
\includegraphics[scale=0.5]{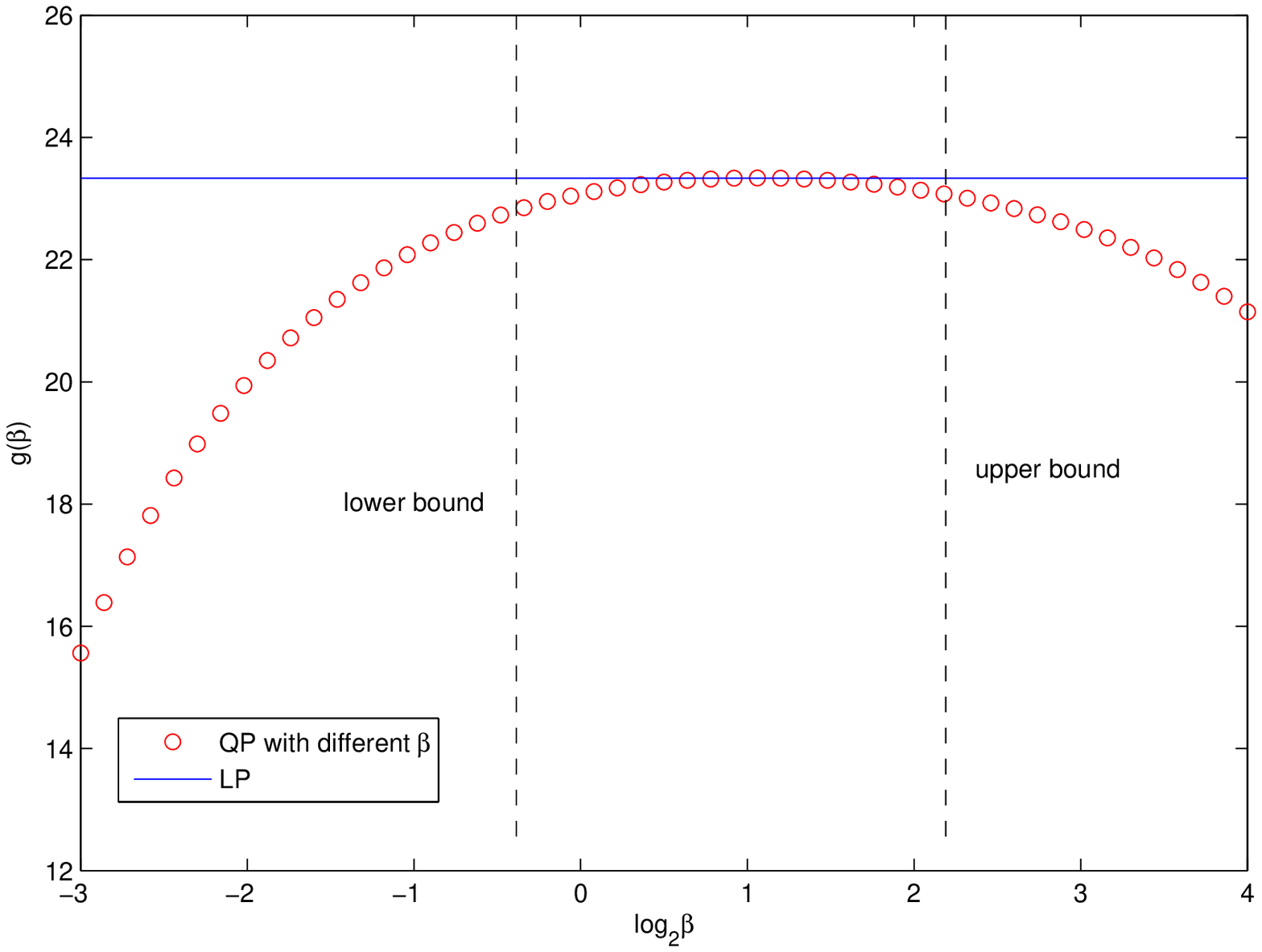}
\caption{Experiment in Figs. 3 and 4 Repeated 100 Times: Averaged Minimum from LP and Averaged Dual Function from QP.}\label{fig:fig5}
\end{figure}

\begin{figure}[htp]
\centering
\includegraphics[scale=0.5]{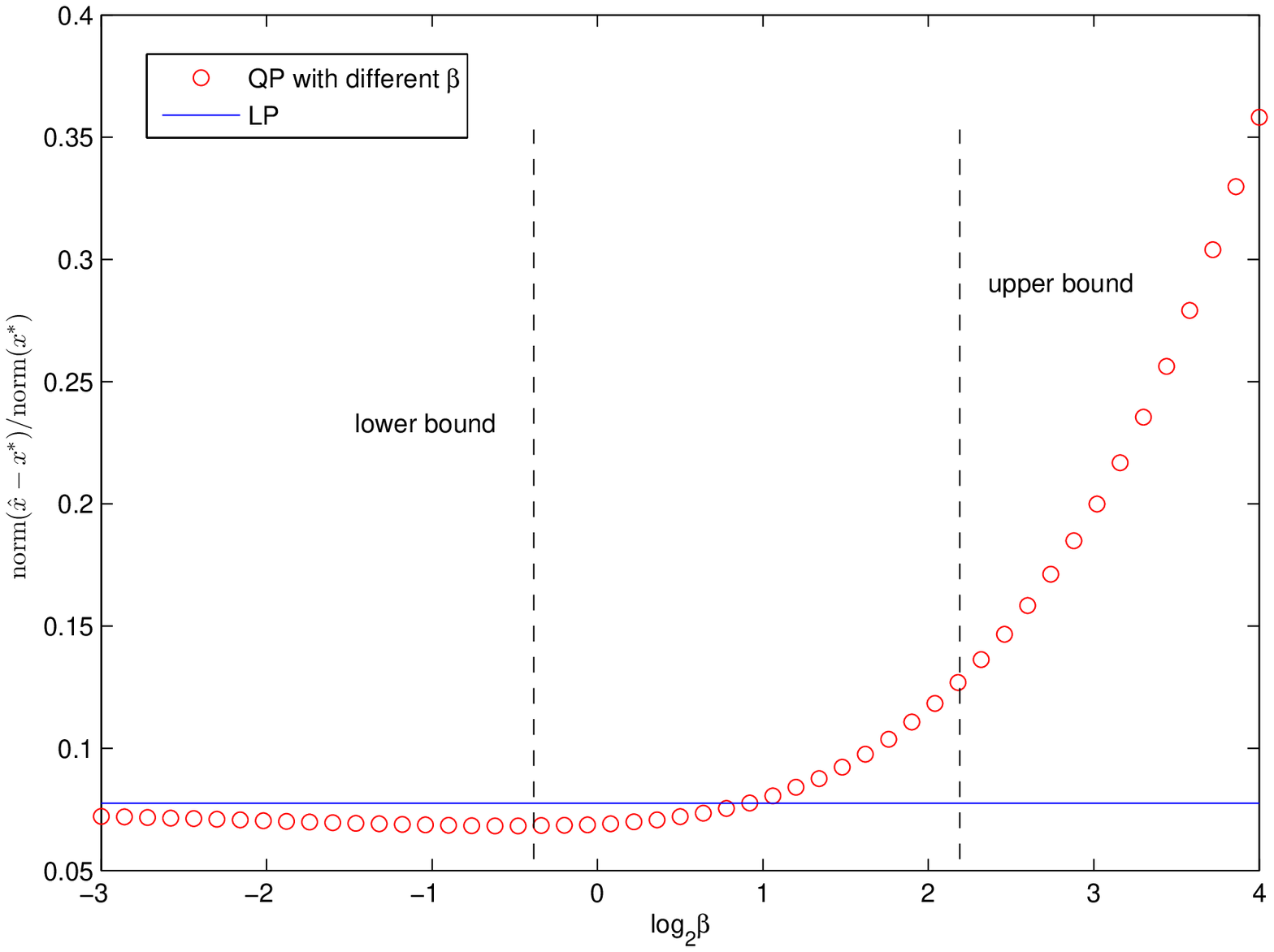}
\caption{Experiment in Figs. 3 and 4 Repeated 100 Times: Averaged Reconstruction Errors from LP and QP.}\label{fig:fig6}
\end{figure}

\begin{figure}[htp]
\centering
\includegraphics[scale=0.5]{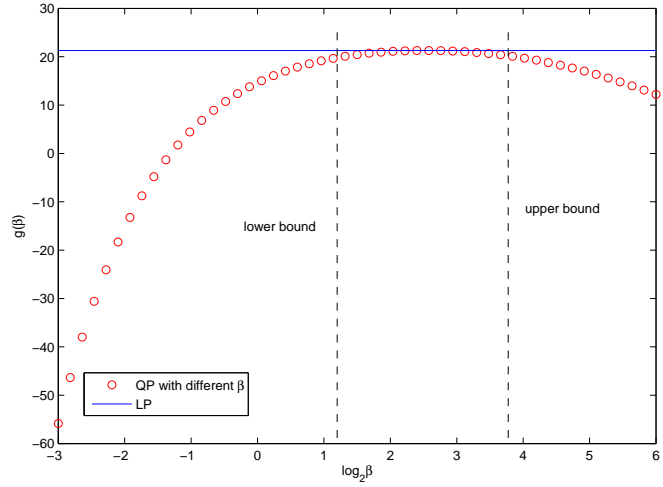}
\caption{The Minimum from LP and the Dual Function $g(\beta)$ from QP: the 20dB SNR Case}\label{fig:fig7}
\end{figure}

\begin{figure}[htp]
\centering
\includegraphics[scale=0.5]{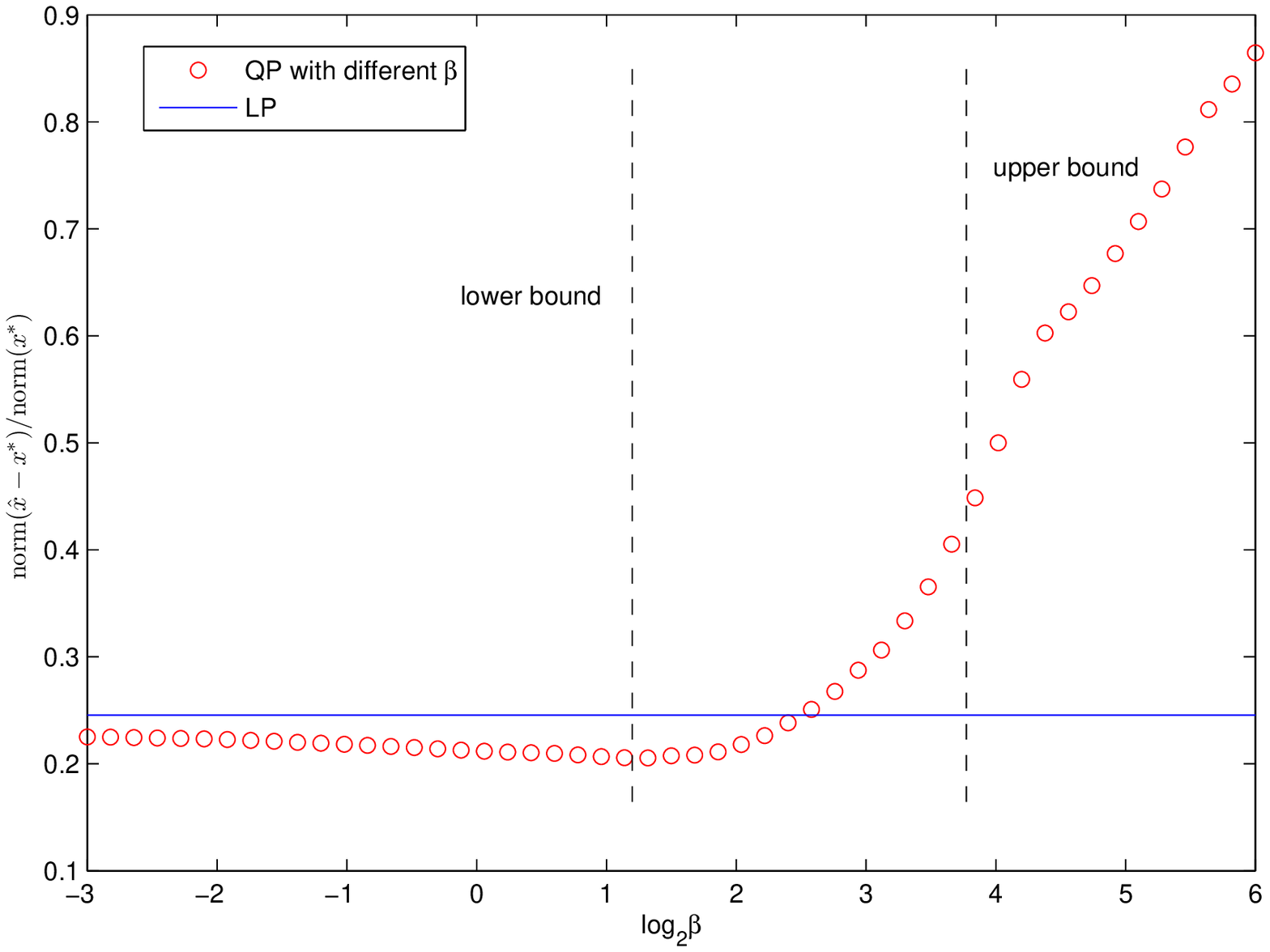}
\caption{Recovery Error from LP and QP: the 20dB SNR Case. The straight line: LP, the curve: QP.}\label{fig:fig8}
\end{figure}

\begin{figure}[htp]
\centering
\includegraphics[scale=0.5]{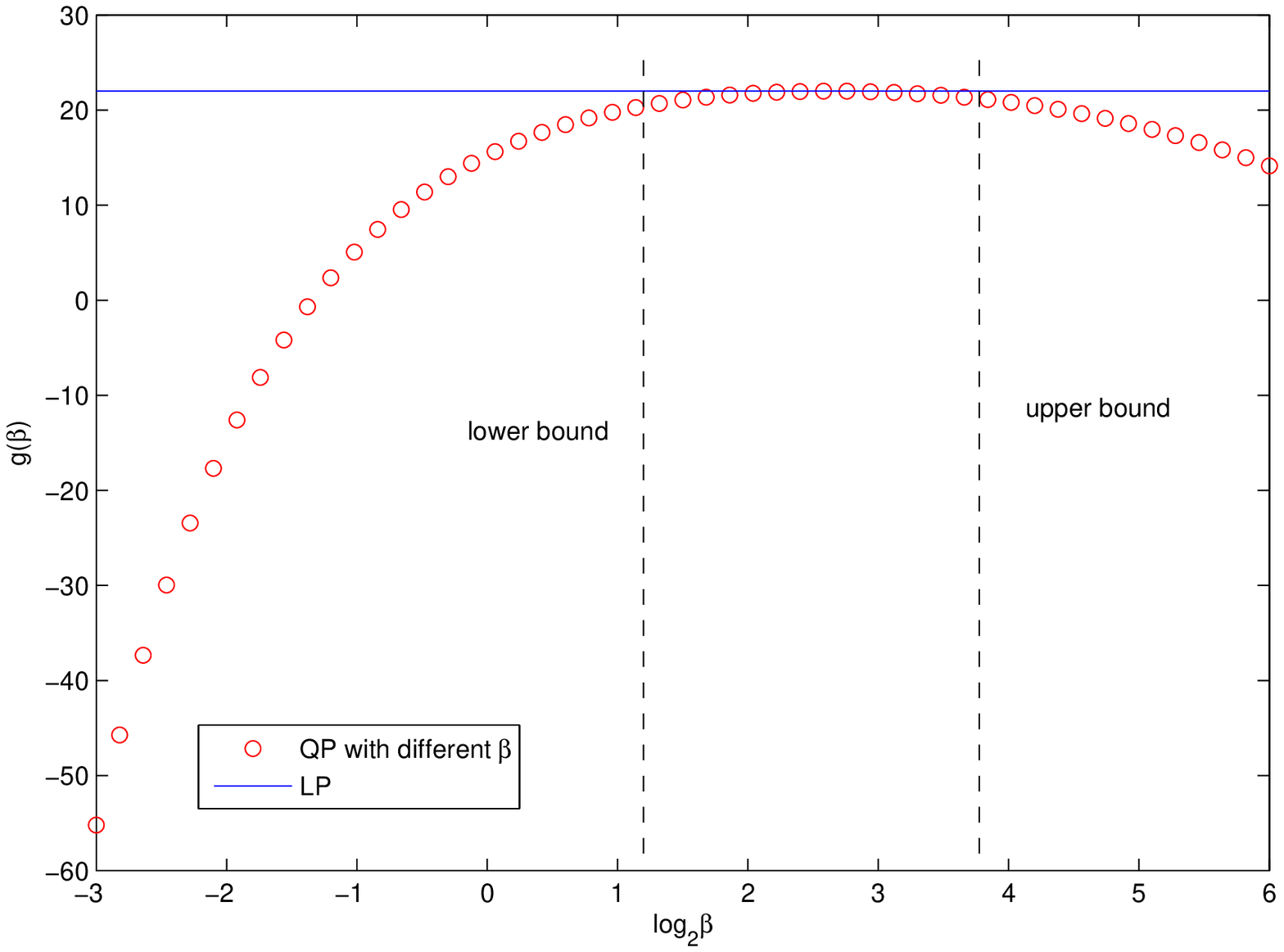}
\caption{Experiment in Figs. 7 and 8 Repeated 100 Times: Averaged Minimum from LP and Averaged Dual Function from QP.}\label{fig:fig13}
\end{figure}

\begin{figure}[htp]
\centering
\includegraphics[scale=0.5]{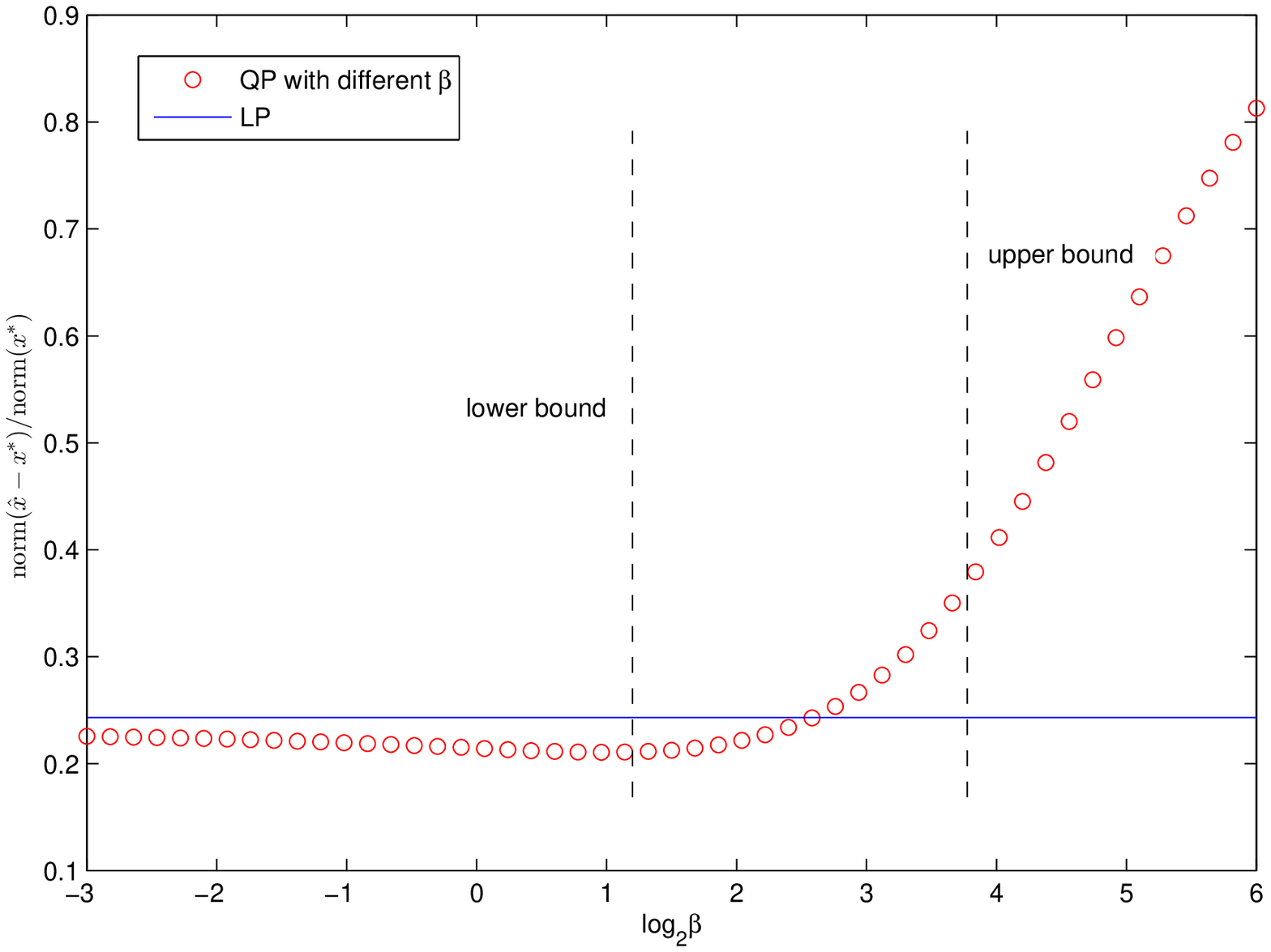}
\caption{Experiment in Figs. 7 and 8 Repeated 100 Times: Averaged Reconstruction Errors from LP and QP.}\label{fig:fig14}
\end{figure}
